\documentclass[fleqn,12pt]{article}
\usepackage{amsmath,latexsym,amssymb,cite}

\newcommand{\bls}[1]{\renewcommand{\baselinestretch}{#1}}
\def\noi{\noindent}

\newcommand{\Author}[2]{\noi{\bf #1}\\[2ex]\noindent{\small\it #2}\\}

\newcommand{\Abstract}[1]{\vskip 2mm \begin{center}
        \parbox{16.4cm}{\small\noi #1} \end{center}\medskip}

\newcommand{\foom}[1]{\protect\footnotemark[#1]}

\newcommand{\email}[2]{\footnotetext[#1]{e-mail: #2}
		\addtocounter{footnote}{1}}

\newcommand{\Theorem}[2]{\medskip\noi {\bf #1. \ }{\sl #2}\medskip}

\def\nqq{\hspace*{-2em}}
\def\nhq{\hspace*{-0.5em}}

\def\cm{\hspace*{1cm}}

\def\Jl#1#2{#1 {\bf #2},\ }

\def\ApJ#1 {\Jl{Astroph. J.}{#1}}
\def\CQG#1 {\Jl{Class. Quantum Grav.}{#1}}
\def\DAN#1 {\Jl{Dokl. AN SSSR}{#1}}
\def\GC#1 {\Jl{Grav. Cosmol.}{#1}}
\def\GRG#1 {\Jl{Gen. Rel. Grav.}{#1}}
\def\JETF#1 {\Jl{Zh. Eksp. Teor. Fiz.}{#1}}
\def\JETP#1 {\Jl{Sov. Phys. JETP}{#1}}
\def\JHEP#1 {\Jl{JHEP}{#1}}
\def\JMP#1 {\Jl{J. Math. Phys.}{#1}}
\def\NPB#1 {\Jl{Nucl. Phys.}{B\ #1}}
\def\NP#1 {\Jl{Nucl. Phys.}{#1}}
\def\PLA#1 {\Jl{Phys. Lett.}{#1A}}
\def\PLB#1 {\Jl{Phys. Lett.}{#1B}}
\def\PRD#1 {\Jl{Phys. Rev.}{D\ #1}}
\def\PRL#1 {\Jl{Phys. Rev. Lett.}{#1}}

\def\al{&\nhq}
\def\lal{&&\nqq {}}
\def\eq{Eq.\,}
\def\eqs{Eqs.\,}
\def\beq{\begin{equation}}
\def\eeq{\end{equation}}
\def\bear{\begin{eqnarray}}
\def\bearr{\begin{eqnarray} \lal}
\def\ear{\end{eqnarray}}
\def\earn{\nonumber \end{eqnarray}}
\def\nn{\nonumber\\ {}}
\def\nnv{\nonumber\\[5pt] {}}
\def\nnn{\nonumber\\ \lal }
\def\nnnv{\nonumber\\[5pt] \lal }
\def\yy{\\[5pt] {}}
\def\yyy{\\[5pt] \lal }
\def\eql{\al =\al}

\def\dst{\displaystyle}
\def\tst{\textstyle}
\def\fracd#1#2{{\dst\frac{#1}{#2}}}
\def\fract#1#2{{\tst\frac{#1}{#2}}}
\def\Half{{\fracd{1}{2}}}
\def\half{{\fract{1}{2}}}
\def\thd{{\fract{1}{3}}}

\def\e{{\,\rm e}}
\def\d{\partial}

\def\sign{\mathop{\rm sign}\nolimits}
\def\diag{\mathop{\rm diag}\nolimits}

\def\const{{\rm const}}
\def\eps{\varepsilon}

\def\then{\ \Rightarrow\ }

\newcommand{\vars}[1]{\left\{\begin{array}{ll}#1\end{array}\right.}
\def\M{{\mathbb M}}
\def\R{{\mathbb R}}
\def\V{{\mathbb V}}
\def\kappa{\varkappa}

\def\Rs{{\mathop{R}\limits_s}{}}
\def\Ro{{\mathop{R}\limits_\omega}{}}
\def\Gs{{\mathop{G}\limits_s}{}}
\def\Go{{\mathop{G}\limits_\omega}{}}
\def\tT{{\widetilde T}{}}

\def\mn{_{\mu\nu}}

\def\mN{_\mu^\nu}

\def\sph{spherically symmetric}

\def\cy{cylindrical}
\def\cyl{cylindrically symmetric}
\def\scyl{static, cylindrically symmetric}

\def\bh{black hole}

\def\wh{wormhole}
\def\whs{wormholes}
\def\asflat{asymptotically flat}

\bls{1.01}
\begin{document}
\thispagestyle{empty}

\section*{\LARGE Rotating cylindrical wormholes}

\bigskip

\Author{\normalsize Kirill A. Bronnikov\foom 1}
{\small Center for Gravitation and Fundam. Metrology, VNIIMS, 46
   Ozyornaya St., Moscow 119361, Russia;\\ Institute of Gravitation and
   Cosmology, PFUR, 6 Miklukho-Maklaya St., Moscow 117198, Russia}

\Author{\normalsize Vladimir G. Krechet}
{\small Yaroslavl State University,
   107 Respublikanskaya St., Yaroslavl 150000, Russia}

\Author{\normalsize Jos\'e P. S. Lemos\foom 2}
{\small Centro Multidisciplinar de {}Astrof\'{\i}sica - CENTRA,
   Departamento de F\'{\i}sica, Instituto Superior T\'ecnico - IST,
   Universidade T\'ecnica de Lisboa - UTL, Avenida Rovisco Pais 1, 1049-001
   Lisboa, Portugal}

\bigskip

\noi {\small\bf Abstract}

\Abstract
  {We consider stationary, \cyl\ configurations in general relativity and
  formulate necessary conditions for the existence of rotating \cy\
  \whs. It is shown that in a comoving reference frame the rotational part
  of the gravitational field is separated from its static part and forms an
  effective stress-energy tensor with exotic properties, which favors the
  existence of \wh\ throats. Exact vacuum and scalar-vacuum solutions (with
  a massless scalar) are considered as examples, and it turns out that even
  vacuum solutions can be of \wh\ nature. However, solutions obtainable in
  this manner cannot have well-behaved asymptotic regions, which excludes the
  existence of \wh\ entrances appearing as local objects in our Universe. To
  overcome this difficulty, we try to build configurations with flat
  asymptotic regions by the cut-and-paste procedure: on both sides of the
  throat, a \wh\ solution is matched to a properly chosen region of flat
  space at some surfaces $\Sigma_-$ and $\Sigma_+$. It is shown, however,
  that if we describe the throat region with vacuum or scalar-vacuum
  solutions, one or both thin shells appearing on $\Sigma_-$ and $\Sigma_+$
  inevitably violate the null energy condition. In other words, although
  rotating \wh\ solutions are easily found without exotic matter, such
  matter is still necessary for obtaining asymptotic flatness.  }

\email 1 {kb20@yandex.ru}
\email 2 {lemos@fisica.ist.utl.pt}

\section{Introduction}

  Wormholes, a subject of active discussion in the current literature, are
  hypothetic objects where two large or infinite regions of space-time are
  connected by a kind of tunnel. These two regions may lie in the same
  universe or even in different universes. The existence of sufficiently
  stable (traversable, Lorentzian) wormholes can lead to physical effects of
  utmost interest, such as a possibility of realizing time machines or
  shortcuts between distant regions of space
  \cite{thorne,viss-book,my_book}. If \whs\ exist on astrophysical scales of
  lengths and times, they can lead to quite a number of unusual observable
  effects \cite{sha, astro}.

  It is well known that a \wh\ geometry can only appear as a solution to the
  Einstein equations if the stress-energy tensor (SET) of matter violates
  the null energy condition (NEC) at least in a neighborhood of the \wh\
  throat \cite{thorne,viss-book,HV97,my_book}. This conclusion, however, has
  been proved under the assumption that the throat is a compact 2D surface,
  having a finite (minimum) area (at least in the static case, since for
  dynamic \whs\ it has proved to be necessary to generalize the notion of a
  throat \cite{HV98}; see, however \cite{HAY98} for other definitions of a
  throat). In other words, it was assumed that, as seen from outside, a
  \wh\ entrance is a local object like a star or a \bh.

  But, in addition to such objects, the Universe may contain structures
  which are infinitely extended along a certain direction, like cosmic
  strings. And, while starlike structures are, in the simplest case,
  described in the framework of spherical symmetry, the simplest stringlike
  configurations are \cyl.

  The opportunity of building \wh\ models in the framework of cylindrical
  symmetry was recently discussed in \cite{cyl-1}. It has been shown, for
  the case of static configurations, that the necessary conditions for the
  existence of \wh\ solutions differs from their \sph\ counterparts, but it
  is still rather hard to obtain more or less realistic \cyl\ \wh\ models.
  Indeed, quite a number of such \wh\ solutions have been obtained, but none
  of them have two flat (or string, i.e., flat up to an angular deficit)
  asymptotic regions, which exclude the existence of \wh\ entrances
  appearing as local objects in our Universe. Moreover, it has been shown
  that the existence of a \scyl, twice \asflat\ \wh\ requires a matter
  source with negative energy density \cite{cyl-1}.

  In this paper we extend the consideration to stationary configurations
  containing a vortex gravitational field. Such fields can lead
  to effective stress-energy tensors with rather exotic properties
  \cite{kr1, kr2, kr3}, which give us a hope to obtain realistic \wh\ models
  in the framework of general relativity.

  A vortex gravitational field is described by the 4-dimensional curl of the
  tetrad $e^\mu_a$: its kinematic characteristic is the angular velocity
  of tetrad rotation
\beq                                                          \label{def-o}
      \omega^\mu = \Half \eps^{\mu\nu\rho\sigma} e_{m\nu} e^m_{\rho;\sigma},
\eeq
  where Greek indices correspond to the four world coordinates $x^\mu$ while
  the Latin letters $m, n, \ldots$ are used for Lorentz indices. The vector
  $\omega^\mu$ determines the effective angular momentum density of the
  gravitational field
\beq
      {}^g S^\mu = \omega^\mu/\kappa,
\eeq
  where $\kappa = 8\pi G$ is Einstein's gravitational constant.

  The simplest example of a space-time possessing a stationary vortex
  gravitational field is that with the \cyl\ metric
\beq                                                            \label{ds0}
     ds^2 = A\,du^2 + C\,dz^2 + B\, d\varphi^2 + 2E\,dt\,d\varphi - D\,dt^2,
\eeq
  where all metric coefficients depend on the ``radial'' coordinate $u$
  whose range is not {\it a priori\/} specified; $z\in \R$ and $\varphi \in
  [0, 2\pi]$ are the longitudinal and azimuthal coordinates, respectively.
  The geometric properties of the spatial sections $t= \const$ of the
  space-time (\ref{ds0}) are determined by the 3D metric
\beq                                                          \label{dl}
       dl^2 = A\,du^2 + C\, dz^2 + \frac{BD + E^2}{D} d\varphi^2.
\eeq

  A vortex gravitational field can both be free (i.e., exist without a
  matter source) and be created by certain fields with polarized spin, such
  as a spinor field; it can certainly exist in the presence of matter
  without polarized spin, such as a perfect fluid. Some examples of
  \cyl\ \wh\ solutions with such sources and the metric (\ref{ds0}) have
  been obtained in \cite{kr1, kr2, kr3}.

  In the present paper, we will discuss the general conditions for the
  existence of \cy\ \whs\ with the stationary metric (\ref{ds0}) with
  rotation and try to obtain a \wh\ model with two flat asymptotic regions
  without invoking exotic matter. Section 2 is devoted to working out some
  general relations for the metric (\ref{ds0}). It is shown that in a
  reference frame comoving to the matter source of gravity (that is, an
  azimuthal matter flow is absent), then the rotational part of the Ricci and
  Einstein tensors can be separated from its ``static'' part, which makes
  much easier the subsequent analysis. Even more than that, the rotational
  part of the Einstein tensor can be viewed as an addition to the stress
  tensor of matter, and the unusual properties of this addition can
  hopefully provide \wh\ construction.

  Further on, in Section 3, we discuss the possible definitions of \cy\ \wh\
  throats by analogy with \cite{cyl-1}. With these definitions, we find the
  necessary conditions for the throat existence, which generalize those
  formulated in \cite{cyl-1} for the static case and are actually easier to
  satisfy due to the rotational contribution.

  As the simplest example, in Section 4 we consider exact rotating \wh\
  solutions which exist in vacuum (due to a vortex gravitational field) or
  in the presence of a massless scalar field.

  It turns out that rotating \wh\ solutions obtainable in the present
  framework cannot be \asflat\ in principle because the angular velocity
  does not vanish at large radii. It is therefore suggested to construct
  \asflat\ configurations by the cut-and-paste procedure: on each side of
  the throat, a \wh\ solution can be cut and matched with a properly chosen
  region of flat space, which should be taken in a rotating reference frame
  to allow matching. The inevitable jump in the extrinsic curvature tensor
  of the junction surfaces $\Sigma_+$ and $\Sigma_-$ corresponds to the
  existence of thin shells of matter whose surface SET $S_{ab}$ can be
  calculated in the well-known Darmois-Israel formalism. It is of utmost
  interest whether or not $S_{ab}$ can respect the weak and null energy
  conditions (WEC and NEC, respectively).

  This general procedure is discussed in Section 5, while Section 6
  describes an attempt to construct such a compound \wh\ model using the
  previously obtained vacuum and scalar-vacuum solutions for the inner
  region containing the throat. It is shown that one or both SETs of the
  thin shells on $\Sigma_+$ and $\Sigma_-$ inevitably violate the NEC. This
  means that a twice \asflat\ \wh\ model cannot be constructed without
  exotic matter in the framework under consideration, at least with the aid
  of vacuum or scalar-vacuum solutions. So, even though rotation strongly
  favors the existence of \cy\ \wh\ throats, the problem of building a
  ``realistic'' \wh\ model remains unsolved.

  Section 7 is a brief conclusion, and in the Appendix we outline the ways
  of obtaining exact rotating \cyl\ solutions for scalar fields with nonzero
  self-interaction potentials, postponing their detailed analysis for the
  future.

\section{Basic relations for stationary rotating \cyl\ space-times}

  Let us rewrite the metric (\ref{ds0}) in a
  slightly different notation,
  singling out the three-dimensional linear element (\ref{dl}):
\bear                                                 \label{ds1}
      ds^2 \eql A\,du^2 + C\,dz^2 + r^2 d\varphi^2 - D[dt- (E/D) d\varphi]^2
\yy                                                   \label{ds2}
          \eql \e^{2\alpha}du^2 + \e^{2\mu}dz^2 + \e^{2\beta}d\varphi^2
		  - \e^{2\gamma}(dt - E\e^{-2\gamma}\, d\varphi)^2,
\ear
  where $A,B,C,D,E$ as well as $\alpha,\beta,\gamma,\mu$ are functions of
  the radial coordinate $x^1 =u$, $x^4 = t$ is the time coordinate, and the
  different notations (given here to facilitate comparison with different
  references) are related by
\bearr
   \begin{array}{lll}
	A = \e^{2\alpha}, \cm & C = \e^{2\mu}, \cm & D = \e^{2\gamma},
   \yy
	\e^{\beta} = r(u), & r^2 = \Delta/D, & \Delta = BD + E^2.
   \end{array}
\ear
  The absolute value of the metric determinant is
\beq
     g = |\det(g\mn)|=  AC\Delta = \e^{2\alpha + 2\beta + 2\gamma + 2\mu}.
\eeq

  In the gauge $A = C$ (equivalently, $\alpha = \mu$) the vortex $\omega
  = \sqrt{\omega_\alpha \omega^\alpha}$ is \cite{kr3,kr09}
\beq                                                  \label{o-def}
     \omega = \frac{E'D - ED'}{2D \sqrt{A\Delta}}
     	    = \Half (E\e^{-2\gamma})' \e^{\gamma-\beta-\mu}.
\eeq

  The Ricci tensor components in the gauge $\alpha = \mu$ can be expressed
  as follows (see also \cite{santos}; the prime stands for $d/du$):
\bear     	\label{r11}
	R^1_1 \eql -\e^{-2\mu}[\beta''+\gamma''+\mu''
		+ \beta'^2 + \gamma'^2 + \mu'(\beta'+\gamma')] + 2\omega^2;
\yy        	\label{r22}
	R^2_2 \eql -\e^{-2\mu}[\mu'' + \mu'(\beta'+\gamma')];
\yy        \label{r33}
	\sqrt{g} R^3_3 \eql -\biggl[\frac{DB' + EE'}{2\sqrt{\Delta}}\biggr]'
	         	       = -\biggl[
	       	       \beta'\e^{\beta+\gamma} - E\omega\e^\mu\biggr]';
\yy        	\label{r44}
	\sqrt{g} R^4_4 \eql -\biggl[\frac{D'B + EE'}{2\sqrt{\Delta}}\biggr]'
	         	       = -\biggl[
	      	       \gamma'\e^{\beta+\gamma} + E\omega\e^\mu\biggr]';
\yy        	\label{r34}
	\sqrt{g} R^3_4 \eql -\biggl[\frac{DE'-ED'}{2\sqrt{\Delta}}\biggr]'
	    	       = -\biggl(\omega \e^{2\gamma+\mu}\biggr)';
\yy        	\label{r43}
	\sqrt{g} R^4_3 \eql -\biggl[\frac{B'E-BE'}{2\sqrt{\Delta}}\biggr]'.
\ear

  Assuming that our rotating reference frame is comoving to the matter
  source of gravity, that is, the azimuthal flow $T^3_4 =0$,\footnote
  	{It is this component of the stress-energy tensor that
	 describes the flow in the $x^3$ direction, or, more precisely,
	 $T^3_4/\sqrt{|g_{44}|}$, as follows from Zel'manov's prescription
	 for spatially covariant and chronometrically invariant vector
	 components \cite{zelmanov}; this expression can also be verified,
	 though with more effort, using the tetrad formalism, see, e.g.,
	 \cite{MTW}.}
  we find from $R^3_4 =0$ that
\beq       	      					\label{omega}
	\omega = \omega_0 \e^{-\mu-2\gamma}, \cm \omega_0 = \const.
\eeq
  Substituting (\ref{omega}) into the expression $(E \omega\e^\mu)'$,
  taking into account (\ref{o-def}), we find that
\beq                                             \label{om'}
       (E\omega\e^\mu)' = 2\omega_0^2 \e^{\beta-3\gamma}
			= 2\omega^2 \e^{2\mu+\beta+\gamma}
			= 2\omega^2 \sqrt{g}.
\eeq

  As a result, in an arbitrary gauge the diagonal components of the Ricci
  tensor can be written as follows:
\bearr           		           \label{R11}
	R^1_1 = -\e^{-2\alpha}[\beta''+\gamma''+\mu''
		 + \beta'^2 + \gamma'^2 + \mu'^2
		 	-\alpha'(\beta'+\gamma'+\mu')] + 2\omega^2;
\yyy 			                    \label{R22}
	R^2_2 = \Box_1 \mu;
\yyy                    		    \label{R33}
	R^3_3 = \Box_1 \beta + 2\omega^2;
\yyy			                    \label{R44}
	R^4_4 = \Box_1 \gamma - 2\omega^2.
\ear
  where, for any $f(u)$, $\Box_1 f = -g^{-1/2}[\sqrt{g}g^{11}f']'
  = -\e^{-2\alpha}[f'' + f'(\beta'+\gamma'+\mu'-\alpha')]$.
  We see that the diagonal part of the Ricci tensor splits into the static
  part $\Rs\mN$ and the rotational part $\Ro\mN$, where
\beq
	\Ro\mN = \omega^2 \diag (2,\ 0,\ 2,\ -2)
\eeq
  (the coordinates are ordered as follows: $u,z,\varphi,t$). The corresponding
  Einstein tensor $G\mN = R\mN - \half \delta\mN R$ splits in a similar
  manner,
\beq                                                          \label{G_mn}
	G\mN = \Gs\mN + \Go\mN, \cm
	\Go\mN = \omega^2 \diag (1,\ -1,\ 1\ -3).
\eeq
  One can check that the tensors $\Gs\mN$ and $\Go\mN$ (each separately)
  satisfy the ``conservation law'' $\nabla_\alpha G^\alpha_\mu =0$ with
  respect to the static metric (the metric (\ref{ds2}) with $E\equiv 0$).

  The Einstein equations are written as
\beq
  	G\mN = - \kappa T\mN, \cm  \kappa = 8\pi G,            \label{EE1}
\eeq
  where, as usual, $G$ is Newton's constant of gravity and $T\mN$ is
  the stress-energy tensor (SET) of all kinds of matter. Equivalently,
\beq                                                           \label{EE2}
  	R\mN = - \kappa \tT\mN, \cm
			\tT\mN = T\mN - \Half \delta\mN T^\alpha_\alpha.
\eeq

  It is clear that the tensor $\Go\mN$ (divided by $\kappa$) behaves in the
  Einstein equations as an additional SET with very exotic properties
  (for instance, the effective energy density is $-3\omega^2/\kappa < 0$),
  acting in an auxiliary static space-time with the metric (\ref{ds2}) where
  $E\equiv 0$.

  Noteworthy, there remains the off-diagonal component (\ref{r43}) which
  is, in general, nonzero. If we assume, as before, $T^3_4 =0$ (the comoving
  reference frame), then
\beq                                                   \label{r43'}
	R^4_3 = \frac{E}{D} (R^3_3 - R^4_4).
\eeq
  Thus, if the diagonal components of the Einstein equations have been
  solved, the ${4\choose 3}$ component holds automatically and need not be
  considered. The same relation as (\ref{r43'}) holds for the SET
  components:  $T^4_3 = (E/D) (T^3_3 - T^4_4)$, it is thus nonzero if
  $T^3_3\ne T^4_4$.

\section{The \cy\ \wh\ geometry and its existence conditions}

  Let us begin with a definition formulated by analogy with Definition 1 in
  \cite{cyl-1}.

\Theorem{Definition 1}
  {We say that the metric (\ref{ds2}) describes a {\bf \wh} geometry if the
  circular radius $r(u)\equiv \e^{\beta(u)}$ has a minimum $r(u_0) > 0$ at
  some $u=u_0$, such that on both sides of this minimum $r(u)$ grows to much
  larger values than $r(u_0)$, and, in some range of $u$ containing $u_0$,
  all metric functions in (\ref{ds2}) are smooth and finite {\rm (which
  guarantees regularity and absence of horizons)}.

  The cylinder $u=u_0$ is then called a {\bf throat} (or an $r$-throat).
  }

  The notion of a \wh\ is, as in other similar cases, not rigorous because
  of the words ``much larger'', but the notion of a throat as a minimum of
  $r(u)$ is exact.

  Now, let us take the diagonal part of the SET $T\mN$ in the most general
  form
\beq                                                           \label{SET}
      T^1_1 = -p_r, \ \ T^2_2 = -p_z, \ \ T^3_3 = -p_\varphi,
      \ \ T^4_4 = \rho,
\eeq
  where $\rho$ is the density and $p_i$ are pressures of any physical origin
  in the respective directions.

  It is straightforward to find out how the SET components should behave on
  a \wh\ throat. At a minimum of $r(u)$, due to $\beta'=0$ and
  $\beta'' > 0$,\footnote
       {Here and henceforth we restrict ourselves for convenience to generic
	minima, at which $\beta'' > 0$. If there is a special minimum
	at which $\beta'' =0$, we still have $\beta'' >0$ in its vicinity,
	along with all consequences of this inequality. The same concerns
	minima of $a(u)$ discussed below.}
  we have $R^3_3 - 2\omega^2 < 0$, and from the corresponding component of
  (\ref{EE2}) it follows
\beq                                                          \label{thr1}
      \rho - p_r - p_z + p_\varphi- 2\omega^2/\kappa < 0.
\eeq
  As noted in \cite{cyl-1}, in the general case of anisotropic pressures,
  (\ref{thr1}) does not necessarily violate any of the standard energy
  conditions even for non-rotating configurations; it is clear, however,
  that the inequality (\ref{thr1}) cannot hold if $\rho$ is substantially
  larger than any of the pressures; for isotropic (Pascal) fluids, with
  all $p_i = p$, the condition (\ref{thr1}) leads to $\rho < p$ if
  $\omega=0$. With nonzero rotation it is much easier to build a
  configuration with a throat, as is illustrated by numerous examples
  \cite{kr1, kr2, kr3}, and some new ones will be presented below.

  The above definition is, however, not unique: thus, it is rather natural,
  also by analogy with spherical symmetry, to define a throat and a \wh\ in
  terms of the area function $a(u)= \e^{\beta+\mu}$ instead of $r(u)$
  \cite{cyl-1}. We will call it an {\it a-throat\/} for brevity.

\Theorem{Definition 2}
  {In a space-time with the metric (\ref{ds2}), an $a$-{\bf throat} is a
  cylinder $u = u_a$ where the function $a(u) = \e^{\beta+\mu}$ has a
  regular minimum.

  A configuration where on both sides of $u_a$ the function $a(u)$ grows to
  values $a \gg a(u_a)$, and, in some range of $u$ containing $u_a$, all
  metric functions in (\ref{ds2}) are smooth and finite, is called an
  $a$-{\bf \wh}.  }

  Let us look what are the existence conditions of an $a$-throat.
  By Definition 2, at $u=u_1$ we have $\beta'+\mu' =0$ and $\beta''+\mu''
  >0$. The minimum occurs in terms of {\it any\/} admissible coordinate $u$,
  in particular, in terms of the harmonic coordinate (\ref{harm}). Using it
  in \eqs (\ref{EE2}) with (\ref{R22}) and (\ref{R33}), we find that the
  condition $\beta''+\mu''$ implies
\beq                                                    	\label{thr2a}
     R^2_2 + R^3_3 = -\e^{-2\alpha}(\beta''+\mu'')+ 2 \omega^2
     			< 2\omega^2
     			\ \ \then\ \
     T^1_1 + T^4_4 = \rho - p_r < 2\omega^2/\kappa.
\eeq
  In addition, substituting $\beta'+\mu'=0$ into the Einstein equation $G^1_1
  = -\kappa T^1_1$, we find
\beq                                                         \label{thr2b}
     G^1_1 = \e^{-2\alpha}\beta'\mu' + \omega^2
     	= -\e^{-2\alpha}\beta'{}^2 + \omega^2 \leq \omega^2
     \ \ \then \ \    -T^1_1 = p_r \leq \omega^2/\kappa.
\eeq
  For nonrotating configurations ($\omega=0$), the requirements
  (\ref{thr2a}) and (\ref{thr2b}) yield \cite{cyl-1}
\[
	 \rho < p_r \leq 0\ \ \ {\rm at}\ \ u=u_1,
\]
  i.e., there is necessarily a region with negative energy density $\rho$ at
  and near an $a$-throat. With $\omega \ne 0$, these requirements leave an
  opportunity of having a \cy\ \wh\ geometry without violating the standard
  energy conditions.

  This, however, concerned only the existence of throats, leaving aside the
  asymptotic behavior of possible solutions. We only note that if we wish to
  have regular asymptotic behaviors far from the throat, such that, in
  particular, $\mu \to \const$ while $r \to \infty$, then it will be a \wh\
  geometry by both definitions.

\section{Vacuum and massless scalar field solutions}

  Consider a minimally coupled scalar field $\phi$ with the Lagrangian
\beq
	L_s = - \Half \eps (\d\phi)^2 -V(\phi)
\eeq
  as a source of the geometry (\ref{ds2}). Here, $\eps = +1$ corresponds to
  a normal scalar field and $\eps=-1$ to a phantom one.

  Let us assume $\phi = \phi(u)$ and the comoving reference frame, so that
  $T^3_4 =0$, and the Ricci tensor has the form (\ref{R11})--(\ref{R44}).
  The stress-energy tensor of $\phi$ is
\beq                                                          \label{SET1}
       T\mN (\phi) = \Half \e^{-2\alpha}\phi'^2 \diag(-1,-1,-1,\ 1)
			+ \delta\mN V(\phi).
\eeq
  It is convenient to use the Einstein equations in the form
\beq                                                           \label{EE}
      R\mN = -\kappa (T\mN -\half \delta\mN T^\alpha_\alpha)
	   = -\kappa (\eps \d_\mu\phi\, \d^\nu\phi - \delta\mN V).
\eeq

  For further consideration we focus on vacuum configurations ($T\mN \equiv
  0$) and those with a massless scalar field ($V(\phi) \equiv 0$). Some
  solutions can also be obtained with nonzero potentials $V(\phi)$, as will
  be outlined in the Appendix.

  Using the harmonic radial coordinate $u$ such that
\beq
	\alpha = \beta + \gamma + \mu,                	\label{harm}
\eeq
  the expressions (\ref{R11})--(\ref{R44}) are substantially simplified, in
  particular, $\Box_1 f = -\e^{-2\alpha}f''$ for any $f(u)$.  In this case
  three of the four diagonal components of (\ref{EE}) for a massless scalar
  field give
\bearr
       R^2_2=0 \ \then \  \mu'' =0,
\yyy
       R^3_3=0 \ \then \  \beta'' - 2\omega^2 \e^{2\alpha}=0,
\yyy
       R^4_4=0 \ \then \  \gamma'' + 2\omega^2 \e^{2\alpha}=0,
\ear
  whence it follows
\bearr                                                      \label{mu}
       \mu = -mu   \cm[\mbox{with a certain choice of $z$ scale}],
\yyy                                                         \label{b+g}
       \beta + \gamma = 2hu \ \ \ \,
       			[\mbox{with a certain choice of $t$ scale}],
\yyy                                                         \label{b-g}
       \beta'' - \gamma'' = 4\omega_0^2 \e^{2\beta - 2\gamma}.
\ear
  In obtaining (\ref{b-g}), \eq (\ref{omega}) has been taken into account.
  \eq (\ref{b-g}) is a Liouville equation whose solution can be written
  in the form
\beq 								\label{s1}
	\e^{\gamma-\beta} = 2\omega_0 s(k,u), \cm s(k,u) =
	\vars{
	k^{-1} \sinh ku, & \  k > 0,\ \ u \in \R_+;\\
		      u, & \ k=0, \ \ u \in \R_+; \\
	 k^{-1} \sin ku, & \ k<0, \ \ 0 < u < \pi/|k|.
	}
\eeq
  Here, $h, k, m = \const$, and the other three integration constants have
  been suppressed by choosing scales along the $z$ and $t$ axes in (\ref{mu})
  and (\ref{b+g}) and the origin of $u$ (in (\ref{s1})). Now it is
  straightforward to obtain
\bear
       \e^{2\beta} \eql \frac{\e^{2hu}}{2\omega_0 s(k,u)},\cm\ \ \
       \e^{2\mu} = \e^{-2mu},
\nnv                                                      \label{sol1}
       \e^{2\gamma} \eql 2\omega_0 s(k,u) \e^{2hu},\cm
       \e^{2\alpha} = \e^{(4h-2m)u},  \cm
       \omega = \frac{\e^{mu - 2hu}} {2s(k,u)},
\nnv
       E \eql \e^{2hu}s(k,u) \int \frac{du}{s^2(k,u)}
         = \e^{2hu}[E_0 s(k,u) - s'(k,u)], \ \ \ E_0 = \const.
\ear

  The scalar field equation reads $\phi''=0$, whence $\phi=Cu$ (fixing the
  inessential zero point of $\phi$); the constant $C$ has the meaning of
  scalar charge density. Lastly, the Einstein equation $G^1_1 = -\kappa
  T^1_1$, which is first-order, leads to a relation between the integration
  constants:
\beq                                                   \label{int1}
	k^2 \sign k = 4(h^2 - 2hm) - 2\kappa\eps C^2.
\eeq
  This completes the solution.

  A \wh\ geometry by Definition 1 corresponds to the cylindrical radius $r =
  \e^\beta \to \infty$ at both ends of the $u$ range. It is clear that $r\to
  \infty$ and $\e^\gamma \to 0$ as $u\to 0$ in all these solutions. In the
  same limit, the vortex $\omega \to \infty$, which probably indicates a
  singularity, although all components of the scalar field SET (\ref{SET1})
  are finite (since $\alpha$ is finite and $\phi' = C$), hence, by the
  Einstein equations, the same is true for the components of the Ricci
  tensor; however, taken separately, the static and vortex parts of the
  Ricci tensor diverge.

  As to the other end of the $u$ range, the situation is more diverse:

\medskip\noi
{\bf 1. $k > 0$.}
  In this case we have $\e^{2\beta} \sim \e^{(2h-k)u}$ and $\e^{2\gamma} \sim
  \e^{(2h+k)u}$ at large $u$, hence a \wh\ geometry by Definition 1 takes
  place for $0 < k < 2h$; we also have $\e^\gamma \to \infty$ at large $u$.
  Definition 2 requires in a similar way $0 < k < 2(h-m)$.

\medskip\noi
{\bf 2. $k = 0$.}
  In this case we have $\e^{2\beta} \sim u^{-1}\e^{2hu}$ and
  $\e^{2\gamma} \sim u \e^{2hu}$, hence we have a \wh\ geometry by
  Definition 1 as long as $h >0$ and by Definition 2 as long as $h-m > 0$.
  We have again $\e^\gamma \to \infty$ at large $u$.

\medskip\noi
{\bf 3. $k < 0$.}
  It is clear that a \wh\ geometry (by both definitions 1 and 2) is
  described by all solutions with $k < 0$, for which the range of $u$ is $0 <
  u < \pi/|k|$ without loss of generality. At both ends, $\e^\beta \to
  \infty$ and $\e^{\gamma}\to 0$, while $\e^{\beta + \gamma}$ and $\e^\mu$
  remain finite, and $ \omega \sim \e^{-2\gamma} \to \infty$.

  From (\ref{int1}) it is evident that $k < 0$ is compatible
  with any $\eps$, and, curiously, $\eps =+1$ even favors negative $k$,
  unlike similar solutions for the static case \cite{cyl-1}.

  A vacuum solution corresponds to $C = 0$, it has been considered
  previously in \cite{kr2, kr3}. It is clear that its special case must
  correspond to flat space in a rotating reference frame. One can verify
  that this special case is $m = 0$, $k=1$, $h =-1/2$: we then obtain the
  metric (\ref{ds_M}) (see below) transformed to the harmonic coordinate
  $u$.

  The limit $\omega_0 \to 0$ leads to the (scalar-)vacuum solution without
  rotation in the form given in \cite{cyl-1} (corresponding to $\beta''
  =\gamma'' =\mu''=0$); however, the transition is not very simple. It is
  only possible for $k > 0$ and $k > 2h$; it is carried out by making a shift
  $u \mapsto u + u_0$ and then turning $u_0$ to infinity in a special way,
  requiring a finite limit of the expression $(\omega_0/k)\e^{(k-2h)u_0}$;
  then, the scales along the $t$ and $z$ axes must be adjusted.

\section {Asymptotic flatness and thin shells}

  If we wish to describe \whs\ in a flat or weakly curved background
  universe, so that they could be visible to distant observers like
  ourselves, it is necessary to describe them as systems with flat (or
  string) asymptotic behaviors. However, when dealing with \cyl\ systems, it
  is rather hard to obtain such solutions: indeed, even the Levi-Civita
  vacuum solution has such an asymptotic only in the special case where the
  space-time is simply flat.

  We have proved previously \cite{cyl-1} for static \cy\ \whs\ that to have
  such asymptotics on both sides of the throat, it is necessary to invoke
  matter with negative energy density. This result does not depend on
  particular assumptions on the nature of matter. If we want to avoid
  negative densities (and exotic matter in general) in \wh\ building, it is
  reasonable to try rotating configurations. It is, however, still harder to
  obtain \asflat\ solutions with rotation than without it. One of the
  difficulties is connected with the fact that the Minkowski metric, being
  written in a rigidly rotating reference frame, is stationary only inside
  the ``light cylinder'', so in order to reach an asymptotic region, the
  rotation should be differential, with vanishing angular velocity at large
  radii. On the other hand, looking at \eq (\ref{omega}), we see that at a
  hypothetic flat infinity, where $\mu$ and $\gamma$ should tend to
  constants, the quantity $\omega$ does not vanish.

  We have seen that even vacuum configurations with rotation can yield
  \cy\ \wh\ throats; however, none of these \whs\ are \asflat.

  A possible way out is to try to cut a non-\asflat\ \wh\ configuration at
  some cylinders $u_+$ and $u_-$ at different sides of the throat and to
  join it there to properly chosen flat space regions. The junction surfaces
  $u=u_+$ and $u=u_-$ will comprise thin shells with certain surface
  densities and pressures, which can in principle be physically plausible.

  It is clear that, for such a purpose, the flat-space metric should be
  taken in a rotating reference frame. To do so, in the Minkowski metric
  $ds^2 = dx^2 + dz^2 + x^2 d\varphi^2 - dt^2$ we can make the substitution
  $\varphi \to \varphi + \Omega t$ to obtain
\beq                                                          \label{ds_M}
      ds_{\rm M}^2 = dx^2 + dz^2 + x^2 (d\varphi + \Omega dt)^2 - dt^2,
\eeq
  where $\Omega = \const$ is the angular velocity of the reference frame.
  The relevant quantities defined above are
\bear                                                      \label{M-param}
      - g_{44} \eql \e^{2\gamma} =  1 - \Omega^2 x^2,
\cm
      r^2 \equiv \e^{2\beta} = \frac{x^2}{1 - \Omega^2 x^2},
\nn
      E \eql \Omega x^2, \cm\ \	\omega = \frac{\Omega}{1 - \Omega^2 x^2}
\ear
  [we are using the general notations according to (\ref{ds2})]. This metric
  is stationary and suitable for matching with the internal metric at $|x| <
  1/\Omega$, inside the ``light cylinder'' $|x| = 1/\Omega$ at which the
  linear rotational velocity reaches that of light.

  To match two \cyl\ regions at a certain surface $\Sigma:\ u=u_0$, it is
  necessary first of all to identify this surface as it is seen from both
  sides, hence to provide coincidence of the two metrics at $u=u_0$, or
  specifically,
\beq                                                           \label{ju-1}
      [\beta] = 0, \qquad [\mu] = 0, \qquad
      [\gamma] = 0, \qquad [E] =0,
\eeq
  where the square brackets denote, as usual, a discontinuity of a given
  quantity across the surface in question,
\[
	[f] : = f(u_0+0) - f(u_0 -0).
\]
  One should note that in general the two metrics to be matched may be
  written using different choices of the radial coordinate $u$, but it does
  not matter since the quantities present in (\ref{ju-1}) are insensitive to
  the choice of $u$.

  The second step is to determine the material content of the matching
  surface $\Sigma$ according to the Darmois-Israel formalism \cite{israel}:
  in our case of a timelike $\Sigma$, the surface stress-energy
  tensor $S_{ab}$ is given by\footnote
      	{Here and henceforth we use the units where $G=1$, so that
	 $\kappa=8\pi$. }
\def\tK {{\tilde K}{}}
\beq                                                        \label{ju-2}
	S_a^b = - \frac{1}{8\pi} [\tK_a^b], \qquad
			\tK_a^b := K_a^b - \delta_a^b K,
\eeq
  where $K = K_a^a$, $K_a^b$ being the extrinsic curvature of the surface
  $\Sigma$. The latter, in turn, is defined in terms of a covariant
  derivative of the unit normal vector $n_\mu$ of $\Sigma$ drawn in a
  chosen direction: if $\Sigma$ is defined by the equation $f(x^\alpha)=0$
  and is parametrized by the coordinates $\xi^a$, then
\beq                                                        \label{def-K}
	K_{ab} = \frac{\d x^\alpha}{\d\xi^a}
			\frac{dx^\beta}{d\xi^b}	\nabla_\alpha n_\beta
	       = -n^\gamma \left[ \frac{d^2 x^\gamma}{\d\xi^a \d\xi^b}
        + \Gamma^\gamma_{\alpha\beta}
	    \frac{\d x^\alpha}{\d\xi^a} \frac{dx^\beta}{d\xi^b}\right].
\eeq
  In our case, we take the surface $x^1 = u =\const$. Choosing the
  directions of the normal $n^\alpha$ to growing $x^1$ and the natural
  parametrization $\xi^a = x^a$, $a = 2,3,4$, we obtain
\beq                                                        \label{K-gen}
	K_{ab} = - \e^{\alpha(u)} \Gamma^1_{ab}
	       = \Half \e^{-\alpha(u)} \frac{\d g_{ab}}{\d u}.
\eeq
  The indices of $K_{ab}$ are raised by the surface metric tensor $g^{ab}$, 
  the inverse of $g_{ab}$, $a,b = 2,3,4$, namely,
\bear                                                        \label{g_ab}
	(g_{ab}) \eql \begin{pmatrix}
		 \e^{2\mu} &  & \\
		   	   & \e^{2\beta}-E^2\e^{-2\gamma}  & E  \\
		  	   & E                             & -\e^{2\gamma}
		   \end{pmatrix},
\nn
	(g^{ab}) \eql \begin{pmatrix}
	 \e^{-2\mu} & & \\
		    & \e^{-2\beta}  & E\e^{-2\beta-2\gamma}\\
	   & E\e^{-2\beta-2\gamma}  & -e^{-2\gamma}+E^2 \e^{-2\beta-4\gamma}
		   \end{pmatrix}.
\ear
  As a result, we can write down the following expressions for the trace $K$
  and the nonzero components of $\tK_a^b$ in terms of the metric (\ref{ds2}):
\bear                                                       \label{tr_K}
	K \eql \e^{-\alpha}(\beta' + \gamma' + \mu'),
\yy							    \label{tK22}
	\tK_2^2 \eql -\e^{-\alpha}(\beta' + \gamma'),
\yy                                                         \label{tK33}
	\tK_3^3 \eql -\e^{-\alpha}(\mu'+\gamma') - E\omega\e^{-\beta-\gamma},
\yy                                                         \label{tK44}
	\tK_4^4 \eql -\e^{-\alpha}(\beta'+\mu') + E\omega\e^{-\beta-\gamma},
\yy                                                         \label{tK34}
	\tK_4^3 \eql \omega \e^{\gamma-\beta}.
\ear
  One can again notice that all quantities (\ref{tr_K})--(\ref{tK34}) are
  insensitive to the choice of the radial coordinate.

  The expression (\ref{tK34}) is of interest because its discontinuity
  describes the surface matter flow in the $\varphi$ direction. Hence, if we
  assume that the thin shell at a junction is at rest in the reference frame
  in which the ambient space-time is considered, we must put $[\omega] =0$.

\section {An attempt to build an \asflat\ wormhole model}

  Let us try to build a model with the following structure:
\beq                                                         \label{stru}
	\M_-  \ \cup \ \Sigma_- \ \cup\ \V \ \cup \ \Sigma_+ \ \cup\ \M_+,
\eeq
  where $\M_-$ and $\M_+$ are regions of Minkowski space described by the
  metric (\ref{ds_M}), $\V$ is a space-time region described by the
  vacuum (or scalar-vacuum) solution (\ref{sol1})--(\ref{int1}) with a
  vortex gravitational field, while $\Sigma_-$ and $\Sigma_+$ are junction
  surfaces endowed with certain surface densities and tensions.

  Both $r$- and $a$-throats should be located in $\V$, therefore the junction
  surface $\Sigma_-$ is assumed to be located at $u = u_-$ small enough,
  such that, in particular, $r'(u_-) < 0$ while the surface $\Sigma_+$ is
  located at $u = u_+$ such that $r'(u_+) > 0$.

  The surfaces $\Sigma_- \in \V$ and $\Sigma_+ \in \V$ are identified with
  the surfaces $x=x_-$ and $x= x_+$ belonging to $\M_-$ and $\M_+$,
  respectively. For the flat metrics in $\M_+$ and $\M_-$ we must admit
  arbitrariness of scales along the $z$ and $t$ axes to provide matching on
  $\Sigma_\pm$, whereas in $\V$ with the metric (\ref{sol1})--(\ref{int1})
  these scales have already been fixed. The angular frequencies $\Omega =
  \Omega_\pm$ can also be different in $\M_+$ and $\M_-$. So $\M_\pm$ will
  be characterized by the metric coefficients
\bearr                                                       \label{M-pm}
      \e^{2\beta} = \frac{x^2}{1-\Omega^2_\pm x^2}; \cm
      \e^\mu = \e^{\mu_\pm }, \cm
      \e^{2\gamma} = \e^{2\gamma_\pm }(1 - \Omega^2_\pm x^2);
\nnn
      \e^{\alpha} = 1, \cm \cm
      		 \omega = \frac{\Omega_\pm}{1-\Omega^2_\pm x^2},
      \cm	 E = \Omega_\pm x^2 \e^{\gamma_\pm },
\ear
  where $\gamma_\pm , \ \mu_\pm ,\ \Omega_\pm $ are constant parameters
  characterizing the regions $\M_+$ and $\M_-$, respectively. Moreover,
  in $\M_+$ it is natural to put, as usual, $x > 0$, so that the range of
  $x$ is there $x_+ < x < \infty$, but the junction can only be located in
  the stationary region, therefore, $x_+ < 1/\Omega_+$. Unlike that,
  in $\M_-$ we put $x < 0$ to adjust the directions of the normal vector
  $n^\mu$ on both sides of the surface $\Sigma_-$, and we similarly have
  $|x_-| < 1/\Omega_-$.

  Now the task is to choose the surfaces $\Sigma_\pm $, to perform matching
  and to calculate the surface densities and pressures. It should be
  stressed again that it is quite unnecessary to adjust the choice of the
  radial coordinate in different regions since all relevant quantities are
  reparametrization-invariant. One should only bear in mind using \eqs
  (\ref{tK22})--(\ref{tK44}) that the prime means a derivative with respect
  to the radial coordinate used in the corresponding region.

  The matching conditions (\ref{ju-1}) on $\Sigma_\pm$ can be written as
  follows:
\bear                                                  \label{j-mu}
	 - mu \eql \mu_{\pm},
\yy                                                    \label{j-gamma}
	 2\omega_0 \e^{2hu} s(k,u)
	 	\eql \e^{2\gamma_\pm}(1-\Omega^2 x^2),
\yy						       \label{j-beta}
	 \frac{\e^{2hu}}{2\omega_0 s(k,u)}
	 	\eql \frac{x^2}{1 - \Omega^2 x^2},
\ear
  where we have dropped the index `$\pm$' with $u,\ x,\ \Omega$. We do not
  write explicitly the condition $[E] = 0$ but assume that it holds. The
  expression for $E$ in (\ref{sol1}) contains the integration constant
  $E_0$, and its choice makes it easy to provide $[E] =0$ on one of the
  surfaces $\Sigma_+$ or $\Sigma_-$. The same condition on the other surface
  then leads to one more constraint on the system parameters, which,
  however, does not affect our further reasoning.

  If we assume, in addition (though it is not necessary), that the surface
  matter is at rest in our reference frame, we must put $[\omega]=0$ which
  gives
\bear                                                    \label{j-omega}
       \frac{\e^{(m-2h)u}}{2 s(k,u)}
       		= \frac{\Omega}{1 - \Omega^2 x^2}.
\ear

  The conditions (\ref{j-mu}) and (\ref{j-gamma}) fix the constants $\mu_\pm$
  and $\gamma_\pm$ and do not affect other constants. So the only constraint
  connecting the parameters of the internal and external regions is
  (\ref{j-beta}).

  Now, the main question is: Can both surface stress-energy tensors be
  physically plausible and non-exotic under some values of the system
  parameters? A criterion for that is the validity of the WEC
  which includes the requirements
\beq                                                             \label{WEC}
	\frac{S_{44}}{g_{44}} = \sigma \geq 0, \ \ \
		S_{ab}\xi^a \xi^b \geq 0,
\eeq
  where $\xi^a$ is an arbitrary null vector ($\xi^a \xi_a =0$) in
  $\Sigma = \Sigma_\pm$, i.e., the second inequality in (\ref{WEC})
  comprises the NEC. The conditions (\ref{WEC}) are
  equivalent to
\beq                                                            \label{WEC'}
      [\tK_{44}/g_{44}] \leq 0, \ \ \ [K_{ab}\xi^a \xi^b] \leq 0.
\eeq

  Let us choose the following two null vectors on $\Sigma$ in the $z$ and
  $\varphi$ directions:
\bear                                                          \label{xi_1,2}
	\xi^a_{(1)} \eql (\e^{-\mu},\ 0,\ \e^{-\gamma}),
\nnv
    \xi^a_{(2)} \eql (0,\ \e^{-\beta},\ \e^{-\gamma}+ E\e^{-\beta-2\gamma})
\ear
  (the components are enumerated in the order $a =2,3,4$).
  Then the conditions (\ref{WEC'}) read
\bearr
       [\e^{-\alpha}(\beta'+\mu')] \leq 0,
\nnnv
       [\e^{-\alpha}(\mu' - \gamma')] \leq 0,
\nnnv
       [\e^{-\alpha}(\beta'-\gamma') + 2\omega] \leq 0,
\ear
  respectively. Now we can apply these requirements to our configuration
  at both junctions. On $\Sigma_-$ with $x_- < 0$ we obtain
\begin{align}                                                 \label{W-}
       \e^{(m-2h)u}\biggl(-m + h - \frac{s'}{2s}\biggr)
       		+ \frac{1}{|x|(1-\Omega^2 x^2)} &\leq 0,
\yy                                                          \label{N1-}
       \e^{(m-2h)u}\biggl(-m - h -\frac{s'}{2s}\biggr)
       		+ \frac{\Omega^2 |x|}{1-\Omega^2 x^2} &\leq 0,
\yy                                                          \label{N2-}
       \e^{(m-2h)u} \, \frac{(1-s')}{s}
       		+ \frac{(1+\Omega x)^2}{|x|(1-\Omega^2 x^2)} &\leq 0
\end{align}
  (where the label ``minus'' with the symbols $u, x, \Omega$ is omitted),
  and on $\Sigma_+$ with $x > 0$ we have similarly
\begin{align}                                                 \label{W+}
       \e^{(m-2h)u}\biggl(m - h + \frac{s'}{2s}\biggr)
       		+ \frac{1}{x(1-\Omega^2 x^2)} &\leq 0,
\yy                                                          \label{N1+}
       \e^{(m-2h)u}\biggl(m + h + \frac{s'}{2s}\biggr)
       		+ \frac{\Omega^2 x}{1-\Omega^2 x^2} &\leq 0
\yy                                                          \label{N2+}
       \e^{(m-2h)u} \, \frac{(s'-1)}{s}
       		+ \frac{(1+\Omega x)^2}{x(1-\Omega^2 x^2)} &\leq 0,
\end{align}
  (with the ``plus'' label omitted). Here $s$ and $s'=ds/du$ refer to
  the function $s=s(k,u)$ introduced in (\ref{s1}).

  The inequalities (\ref{W-}), (\ref{N1-}), (\ref{W+}), (\ref{N1+}) involve
  all the parameters $m$, $h$ and $k$ and are comparatively hard to explore,
  whereas (\ref{N2-}) and (\ref{N2+}) depend in essence on $k$ only. It
  turns out that the latter two lead to the conclusion that the matter
  content of both $\Sigma_+$ and $\Sigma_-$ cannot satisfy the NEC
  (hence also the WEC).

  Indeed, (\ref{N2-}) can hold only if $1 - s'(k,u) < 0$ at $u=u_-$. But
  $s'(k,u) = \{\cosh k, \ 1,\ \cos|k|u\}$ for $k >0,\ k=0$, and $k < 0$,
  respectively, and only at $k>0$ we have $1-s' < 0$. Thus the NEC for
  $\Sigma_-$ definitely requires $k >0$ in the solution valid in $\V$.

  In a similar way, (\ref{N2+}) can hold only if $1 - s'(k,u) >0$ at
  $u=u_+$, and this is only possible if $k <0$. All this means that whatever
  particular solution (with fixed parameters including $k$) is taken to
  describe the space-time region $\V$, the inequalities (\ref{N2-}) and
  (\ref{N2+}) cannot hold simultaneously. Hence the NEC is violated at least
  on one of the surfaces $\Sigma_+$ and $\Sigma_-$.

\section{Conclusion}

  The main purpose of the present study was to make clear whether or not
  rotating \cy\ \whs\ can be obtained without exotic matter and whether or
  not such \whs\ can be \asflat\footnote
	{Instead of asymptotic flatness, one could consider a ``string''
	behavior at large radii, with a finite angular deficit or excess,
	but it is quite clear that our conclusions would be the same.}
  on both sides of the \wh\ throat.  An answer to the first question is
  ``yes'' because the vortex gravitational field creates an effective SET
  with the exotic structure (\ref{G_mn}). Meanwhile, the probable answer to
  the second question is ``no''.

  It is well known that the nontrivial solution of the Laplace equation in
  flat space has a logarithmic asymptotic behavior ($\sim \ln r$) instead of
  vanishing at large $r$. This simple fact actually extends to solutions of
  the Einstein equations, beginning with the Levi-Civita static vacuum
  solution, and makes it difficult to inscribe \cy\ sources into a weakly
  curved environment. As we have seen, the problem is only enhanced if we
  invoke rotation.

  We have tried to overcome this difficulty by considering a configuration
  that consists of (i) a strong-field region $\V$ with the throat, described
  by the simplest rotating \wh\ solution (in vacuum or with a massless
  scalar field), (ii) rotating thin shells $\Sigma_{\pm}$ placed on both
  sides of $\V$, and (iii) two flat-space regions around the shells. It has
  turned out that the surface SET on $\Sigma_{\pm}$ inevitably violates the
  NEC at least on one of the shells (but maybe on both). Let us stress that
  this result has been obtained without any assumption on the angular
  velocity of the shells: they can be at rest in the rotating reference
  frame in which our system is considered or rotate with respect to it at an
  arbitrary rate.

  Thus a twice \asflat\ rotating \cy\ wormhole cannot be built without
  exotic matter even though rotation favors the formation of throats. This
  result has been obtained for vacuum and scalar-vacuum solutions in $\V$,
  but there is a certain hope that it can change with another kind of
  non-exotic matter filling this internal region.

\section*{Appendix: some solutions with nonzero $V(\phi)$}
\def\theequation{A.\arabic{equation}}
\setcounter{equation}{0}

  We will outline here the way of obtaining exact solutions to the field
  equations for $V(\phi) \ne 0$, in the following cases:
\begin{itemize}		\itemsep 0pt
\item
	An exponential potential, $V(\phi) = V_0 \e^{2\lambda\phi}$,
	$V_0,\ \lambda = \const$.
\item
	An arbitrary potential $V(\phi)$.
\end{itemize}
  In the latter case, solutions can be obtained by the inverse problem
  method by specifying one of the metric functions.

  A detailed analysis of these solutions and their possible usage for \wh\
  construction is postponed to future publications.

  We are dealing with the metric (\ref{ds2}) and \eqs (\ref{EE}), where
  $\phi = \phi(u)$, the Ricci tensor components are given by
  (\ref{R11})--(\ref{R44}) and $\omega = \omega_0 \e^{-\mu - 2\gamma}$
  according to (\ref{omega}).

  Let us use, as in Sec.\,4, the harmonic radial coordinate $u$, such
  that $\alpha = \beta + \gamma + \mu$. We have, as before, $R^3_3 = R^4_4$,
  which leads to the Liouville equation (\ref{b-g}) and finally to
  (\ref{s1}), giving us the difference $\eta := \gamma-\beta$. Furthermore,
  we have $R^3_3 + R^4_4 = 2R^2_2$, whence
\beq                                                      \label{mu''}
       2\mu'' = \beta'' + \gamma''\ \then \
       2\mu = \beta + \gamma + au,\ \ a = \const,
\eeq
  suppressing the second integration constant by choosing the scale along
  the $z$ axis. We also see that $\alpha'' = 3\mu'' =(3/2)(\beta+\gamma)''$.
  Thus all metric coefficients are expressed in terms of $\eta$ and
  $\alpha$, namely,
\bear
       2\beta \eql  \eta + \thd (2\alpha - au),
\nn
       2\gamma \eql -\eta + \thd (2\alpha - au),
\nn
       2\mu \eql \thd (2\alpha + 2au),                 \label{A-ds}
\ear
  and the off-diagonal component $E$ is then obtained by combining
  (\ref{o-def}) and (\ref{omega}) [where \eq (\ref{o-def}) should be
  rewritten in the form $2\omega = (E\e^{-2\gamma})'\e^{\gamma-\beta-\alpha}$,
  allowing for an arbitrary $u$ gauge]:
\beq
       (E\e^{-2\gamma})' = 2\omega_0 \e^{2\beta-2\gamma}.
\eeq

  For the two remaining unknowns $\phi$ and $\alpha$, we can use the scalar
  field equation and the component $R^2_2 = ...$ of the Einstein equations
  that yield
\bearr                                                     \label{e-sca}
	\eps \e^{-2\alpha} \phi'' = dV/d\phi,
\yyy                                                       \label{e-alf}
	\e^{-2\alpha} \alpha'' = -3\kappa V.
\ear

  It is also necessary to write the first-order constraint equation $G^1_1 =
  ...$ which has the form
\beq                                                      \label{G11=}
     \Half \kappa\eps \phi'{}^2 = \kappa V \e^{2\alpha}
	+ \omega_0^2 \e^{\mu + 2\beta}
	+ \beta'\gamma' + \beta'\mu' + \gamma'\mu'.
\eeq
  If a solution is found by integrating the second-order Einstein equations,
  then (\ref{G11=}) verifies the validity of this solution and yields a
  relation among the integration constants.

\medskip\noi
{\bf An exponential potential and a cosmological constant.}
  For the potential $V(\phi) = V_0 \e^{2\lambda\phi}$,
  \eqs (\ref{e-sca}) and (\ref{e-alf}) read
\bearr                                                      \label{phi''}
	\eps \phi'' = 2\lambda V_0 \e^{2\lambda\phi + 2\alpha},
\yyy                                                        \label{al''}
	\alpha'' = -3\kappa V_0 \e^{2\lambda\phi + 2\alpha}
\ear
  Two combinations of \eqs (\ref{phi''}) and (\ref{al''}) are easily solved:
  one of them simply connects $\alpha''$ and $\phi''$ while the other is a
  Liouville equation for a linear combination of $\alpha$ and $\phi$:
\bear
	3\eps\kappa\phi'' + 2\lambda\alpha'' \eql 0,     \label{af-1}
\\                                                       \label{af-2}
	(\alpha + \lambda\phi)'' \eql
		 (2\eps\lambda^2 - 3\kappa)\e^{2(\alpha+\lambda\phi)}.
\ear
  It is then necessary to substitute the solution to (\ref{G11=}) to find a
  relation among the integration constants.

  The solutions are easily found for any values of $\eps$, $V_0$ and
  $\lambda$. In particular, if we restrict ourselves to $\eps = +1$ and $V_0
  > 0$, we have three branches of solutions depending on the sign
  of $2\lambda^2-3\kappa$.

  The case of a nonzero cosmological constant $\Lambda$ is simply the
  special case where the potential is constant: $\lambda =0$,
  $\Lambda = \kappa V_0$.

\medskip\noi
{\bf An arbitrary potential.} It is in general hard to solve \eqs
  (\ref{e-sca}) and (\ref{e-alf}) when $V(\phi)$ is specified; however,
  specifying $\alpha(u)$, we can solve the problem completely. Indeed, for
  given $\alpha$, (\ref{e-alf}) readily gives $V(u)$, while $\phi(u)$ is
  found in terms of the metric coefficients (which are all known by now)
  from the first-order equation (\ref{G11=}).

  It remains to select solutions with physical properties of interest.

\subsection*{Acknowledgments}

  KB acknowledges support from the Funda\c{c}\~{a}o para a Ci\^{e}ncia e a
  Tecnologia of Portugal (FCT) through the project PTDC/FIS/098962/2008
  while staying at CENTRA/IST in Lisbon, and thanks the colleagues at CENTRA
  for kind hospitality; JPSL thanks FCT funding through the project
  PEst-OE/FIS/UI0099/2011.

\end{document}